\documentclass[letter]{aa}

\usepackage{graphicx}
\usepackage{txfonts}
\defcitealias{2017A&A...601A..19G}{vL2017}
\begin{document}

\title{Open star clusters in the Milky Way}

\subtitle{Comparison of photometric and trigonometric distance scales based on \textit{Gaia} TGAS data}

\author{Dana~A.~Kovaleva   \inst{1}   \and
        Anatoly~E.~Piskunov   \inst{1,2} \and
        Nina~V.~Kharchenko \inst{2,3} \and
        Siegfried~R\"{o}ser    \inst{2,4}   \and
        Elena~Schilbach    \inst{2,4}   \and
        Ralf-Dieter~Scholz    \inst{5}   \and
        Sabine~Reffert      \inst{2}   \and
        Steffi~X.~Yen        \inst{2}
        }

\offprints{R.-D.~Scholz}

\institute{
Institute of Astronomy, Russian Academy of Sciences, 48 Pyatnitskaya Str., 109017
Moscow, Russia
\and
Zentrum f\"ur Astronomie der Universit\"at
Heidelberg, Landessternwarte, K\"{o}nigstuhl 12, 69117 Heidelberg, Germany
\and
Main Astronomical Observatory, 27 Academica Zabolotnogo Str., 03143 Kiev,Ukraine
\and
Zentrum f\"ur Astronomie der Universit\"at
Heidelberg, Astronomisches Rechen-Institut, M\"{o}nchhofstra\ss{}e 12-14, 69120 Heidelberg, Germany
\and
Leibniz-Institut f\"ur Astrophysik Potsdam, An der Sternwarte 16, 14482
Potsdam, Germany\\
email: rdscholz@aip.de
}

   \date{Received 15 August 2017; accepted 24 September 2017}

  \abstract
   {The global survey of star clusters in the Milky Way (MWSC) is a comprehensive list of 3061 objects that provides, among other parameters, distances to clusters based on isochrone fitting. The Tycho-Gaia Astrometric Solution (TGAS) catalogue, which is a part of Gaia data release 1 (Gaia DR1), delivers accurate trigonometric parallax measurements for more than 2 million stars, including those in star clusters.}
   {We compare the open cluster photometric distance scale with the measurements given by the trigonometric parallaxes from TGAS to evaluate the consistency between these values.}
   {The average parallaxes of probable cluster members available in TGAS provide the trigonometric distance scale of open clusters, while the photometric scale is given by the distances published in the MWSC. Sixty-four clusters are suited for comparison as they have more than 16 probable members with parallax measurements in TGAS.  We computed the average parallaxes of the probable members and compared these to the photometric parallaxes derived within the MWSC.}
   { We find a good agreement between the trigonometric TGAS-based and the photometric MWSC-based distance scales of open clusters, which for distances less than 2.3\,kpc coincide at a level of about 0.1\,mas with no dependence on the distance. If at all, there is a slight systematic offset along the Galactic equator between $30^\circ$ and $160^\circ$ galactic longitude.
  }
   {}


   \maketitle
%

\section{Introduction}
 
 Open star clusters are important building blocks of the Milky Way. The knowledge of their distances is essential for studies of the structure, dynamics, and evolution of the population in the Galactic disk. The most direct method of cluster distance estimation is based on measurements of trigonometric parallaxes of cluster stars. Until now, the astrometric method has been restricted to a few clusters in the solar vicinity. For more remote clusters  distances are usually determined from photometric data by fitting theoretical isochrones to the observed cluster colour magnitude diagrams. One expects the two methods to provide consistent results.
 
A sample of clusters distributed within a significant range of distances determined by the photometric and
astrometric distance scales enables a comparison between the two methods. Within the project Milky Way Star Clusters \citep[MWSC hereafter;][]{2012A&A...543A.156K}, cluster parameters for 2859 clusters known in the literature \citep{2013A&A...558A..53K} and 202 newly discovered open clusters and associations \citep{2014A&A...568A..51S, 2015A&A...581A..39S} were determined. The sample includes clusters with distances up to 15\,kpc with the mode at 2.4\,kpc. The distances in the MWSC survey are based on photometric data of cluster members from the 2MASS catalogue \citep{cat2MASS} that were fitted to isochrones derived from evolutionary stellar models. 

On the other hand, the recently published first Gaia data release, Gaia DR1 \citep{2017A&A...599A..50A}, includes the TGAS catalogue \citep{2016A&A...595A...1G} with parallaxes and proper motions based on a combination of Tycho-2 \citep{tycho2} and Gaia DR1 observations. The TGAS catalogue \citep{2015A&A...574A.115M} presents trigonometric parallaxes of about 2 mln. stars with a nominal accuracy of 0.3\dots0.7\,mas. These are sufficient data to obtain independent average trigonometric parallaxes of some MWSC clusters up to distances of about 2\,kpc. 

This paper is organised as follows: in Section 2, we describe the input data, which were selected from the MWSC survey and cross-identified with TGAS, for our study. The results of the comparison of the photometric and astrometric distance scales for open clusters are discussed in Section 3, and the summary and conclusions are given in Section 4. 

\section{The data}\label{sec:data}

   \begin{figure}
   \centering
   \includegraphics[width=\hsize]{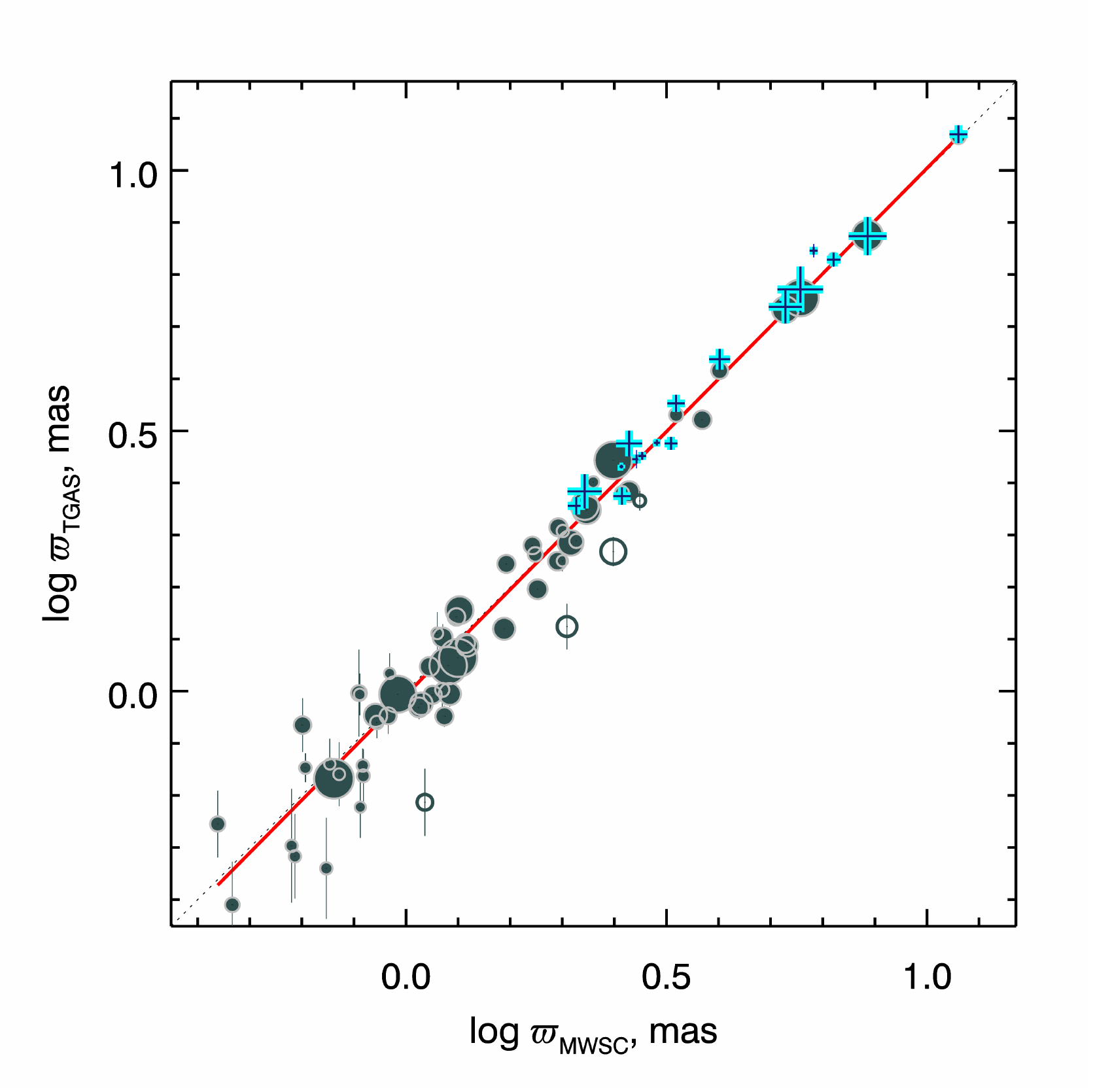}
      \caption{Logarithm of the average trigonometric parallax of clusters $\log \varpi_{TGAS}$ obtained from individual parallaxes of cluster members vs. logarithm of their photometric parallaxes $\log \varpi_{MWSC}$. The sizes of the circles are proportional to the number of TGAS parallaxes available for the calculation of the mean parallax; the vertical bars represent the mean error of the mean. The open circles indicate four discordant clusters discussed in the text. The pluses represent the TGAS parallaxes for selected nearby clusters from \citetalias{2017A&A...601A..19G}, which are shown here for comparison. The red line indicates a linear regression computed. The dotted line indicates the bisector.}
         \label{fig:avplx}
   \end{figure}

The MWSC survey provides a comprehensive sample of star clusters in our Galaxy together with a number of well-determined parameters based on uniform photometric and  kinematic stellar data gathered from the all-sky catalogues 2MASS \citep{cat2MASS} and PPMXL \citep{2010AJ....139.2440R}. The full sample contains 3208 objects: 3061 open and 147 globular clusters. In this study, we concentrate on the subset of open clusters.
For all stars within the cluster areas, membership probabilities were determined by the use of kinematic (proper motions) and photometric (colour magnitude diagrams) selection criteria. The procedure is described in \citet{2012A&A...543A.156K}, and the results are published in the MWSC catalogue of stars in cluster areas  \citep{2013A&A...558A..53K,2014A&A...568A..51S,2015A&A...581A..39S}.
Membership probabilities were used to determine basic parameters of a cluster, in particular its apparent size, proper motion, age, distance, and reddening. For the purpose of this study, we selected stars with membership probability $P>60\%$ and hereafter we refer to these stars as {\it probable} cluster members.

For the cross-match between the {\it probable} cluster members and the stars in Gaia DR1/TGAS, we used the Sky Algorithm by TopCat \citep{2005ASPC..347...29T}.
The limiting angular distance between matched stars was set to be 1 arcsec, although actually it could have been replaced with any value between $0.4~$arcsec and $5~$arcsec without a noticeable change of results. We found that 5743 {\it probable}  members in 1118 MWSC clusters have TGAS parallaxes.  Out of these stars, 199 stars have negative TGAS parallaxes ($\varpi_{TGAS} \leq 0$). The top 10 of the most populated clusters contain more than 50 {\it probable} members with TGAS parallaxes each. As the parallax errors of the {\it probable} members of a given cluster were similar to each other, the average trigonometric parallax of a cluster was computed as the simple, unweighted mean with $3\sigma$ clipping of the outliers. 
We also analysed the MWSC-TGAS proper motion differences 
($\Delta\mu_{RA}, \Delta\mu_{DE}$)  of {\it probable} cluster members.
These did not show systematic effects. Their mean values of
($-0.05\pm0.04, -0.06\pm0.03$)\,mas/yr
were not significant; their standard deviations of (3.04, 2.42)\,mas/yr
reflected the MWSC proper motion accuracies. Only a few outliers were 
found. Therefore, $3\sigma$ clipping 
was also performed on the TGAS proper motions to remove kinematic non-members. Having done this, 
we took all remaining {\it probable} cluster members for the calculation, 
including those with negative parallaxes in TGAS, to avoid a bias in 
the result.
 
Hereafter we refer to the means as $\varpi_{TGAS}$ and to their computed errors as $\varepsilon_\varpi$. For each cluster, the photometric parallax was computed as $\varpi_{MWSC}=1000/d_{MWSC}$, where $\varpi_{MWSC}$ is in mas, and the photometric distance from the MWSC survey  $d_{MWSC}$ is given in parsecs. We determined MWSC distances by fitting cluster member CMDs to isochrones that were computed using the online server CMD2.2\footnote{http://stev.oapd.inaf.it/cgi-bin/cmd} \citep[for more details see][]{2012A&A...543A.156K,2013A&A...558A..53K}.

   \begin{figure}
   \includegraphics[width=0.880\hsize]{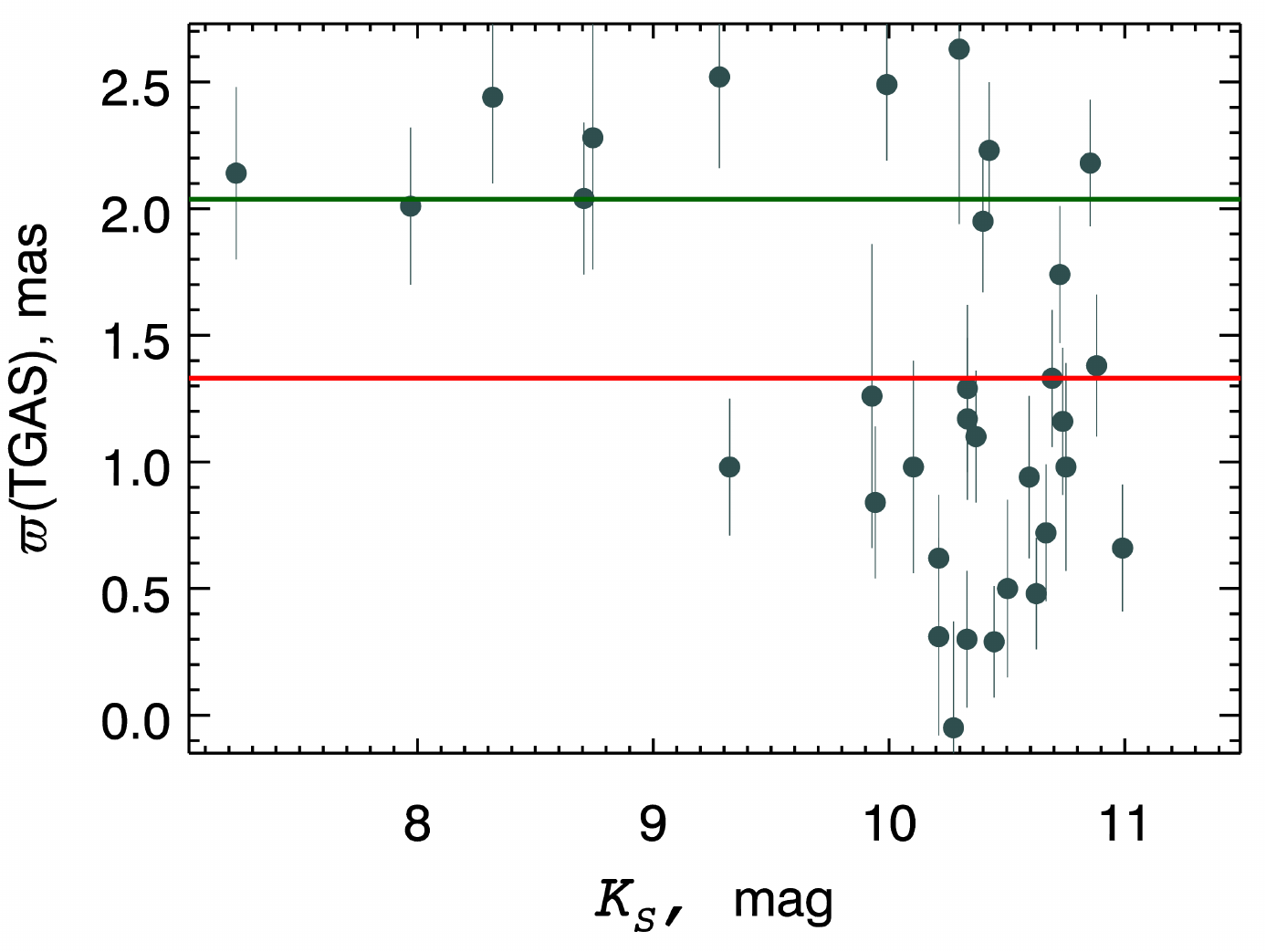}
     \caption{Distribution of the parallaxes of {\it probable} members of the nearby cluster \object{Platais 12} vs. apparent magnitude $K_S$. The bars correspond to the parallax errors. The horizontal lines show the cluster average trigonometric (red) and the photometric (green) parallaxes.}\label{fig:plxks}
    \end{figure}

\section{Trigonometric versus photometric parallaxes}
\label{sec:CDB}
Although we found {\it probable} members with TGAS parallaxes in 1118 MWSC clusters, not all clusters can be used for a comparison between the photometric and trigonometric scales but only those with a certain minimum of {\it probable} members in TGAS. This number should be large enough to diminish the impact of individual stars on the mean trigonometric parallax due to possible uncertainties of their  membership determination and/or  parallax measurements. On the other hand, because of its shallow magnitude limit, as a rule the TGAS catalogue only contains the tips of the cluster CMDs with a few of the brightest stars.
Therefore, as a compromise, for the comparison we used only those clusters for which parallaxes were measured for more than 16 {\it probable} members. As a consequence, the comparison sample included 64 clusters.

   \begin{figure*}
   \includegraphics[width=0.33\hsize]{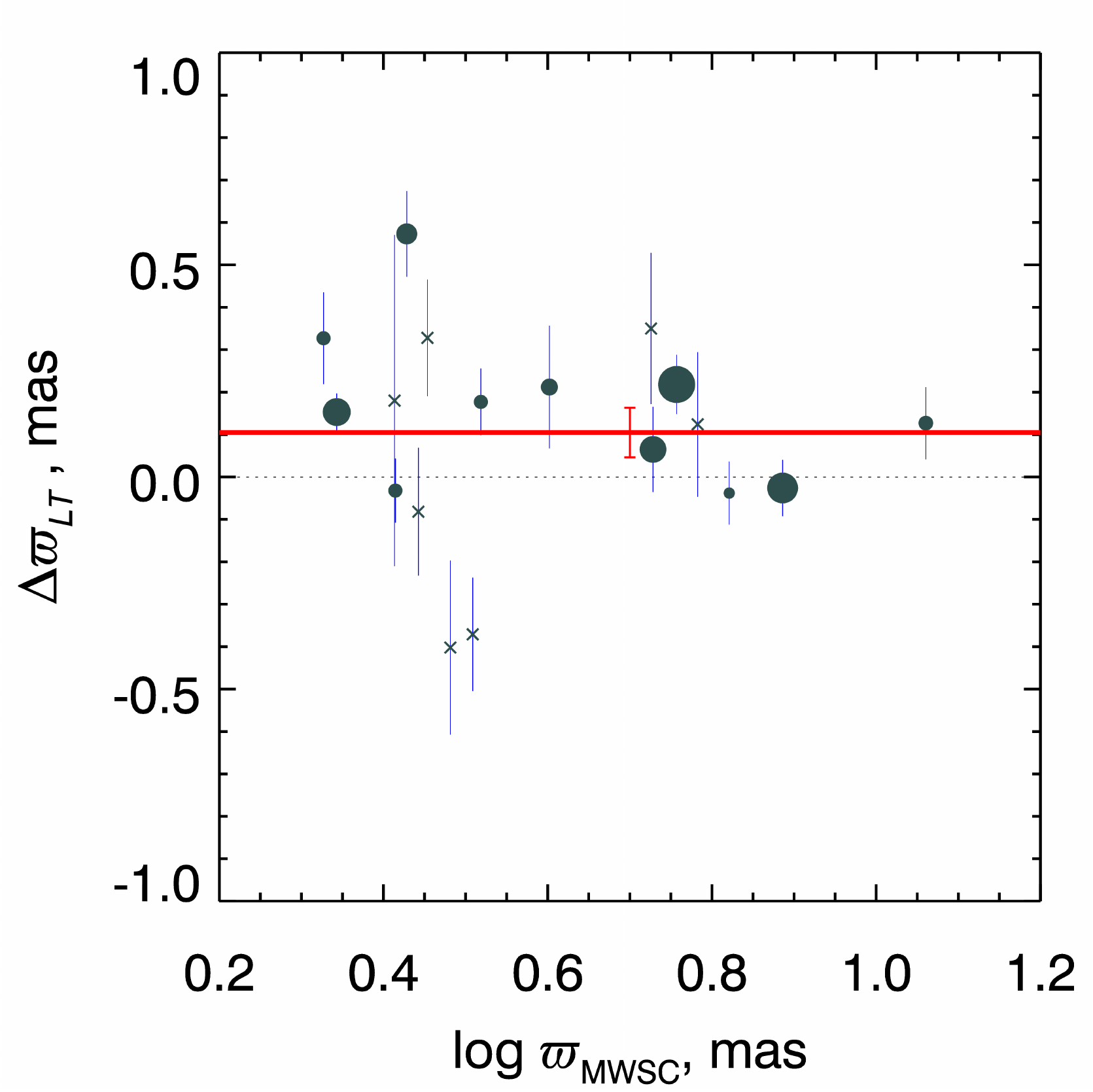}
   \includegraphics[width=0.33\hsize]{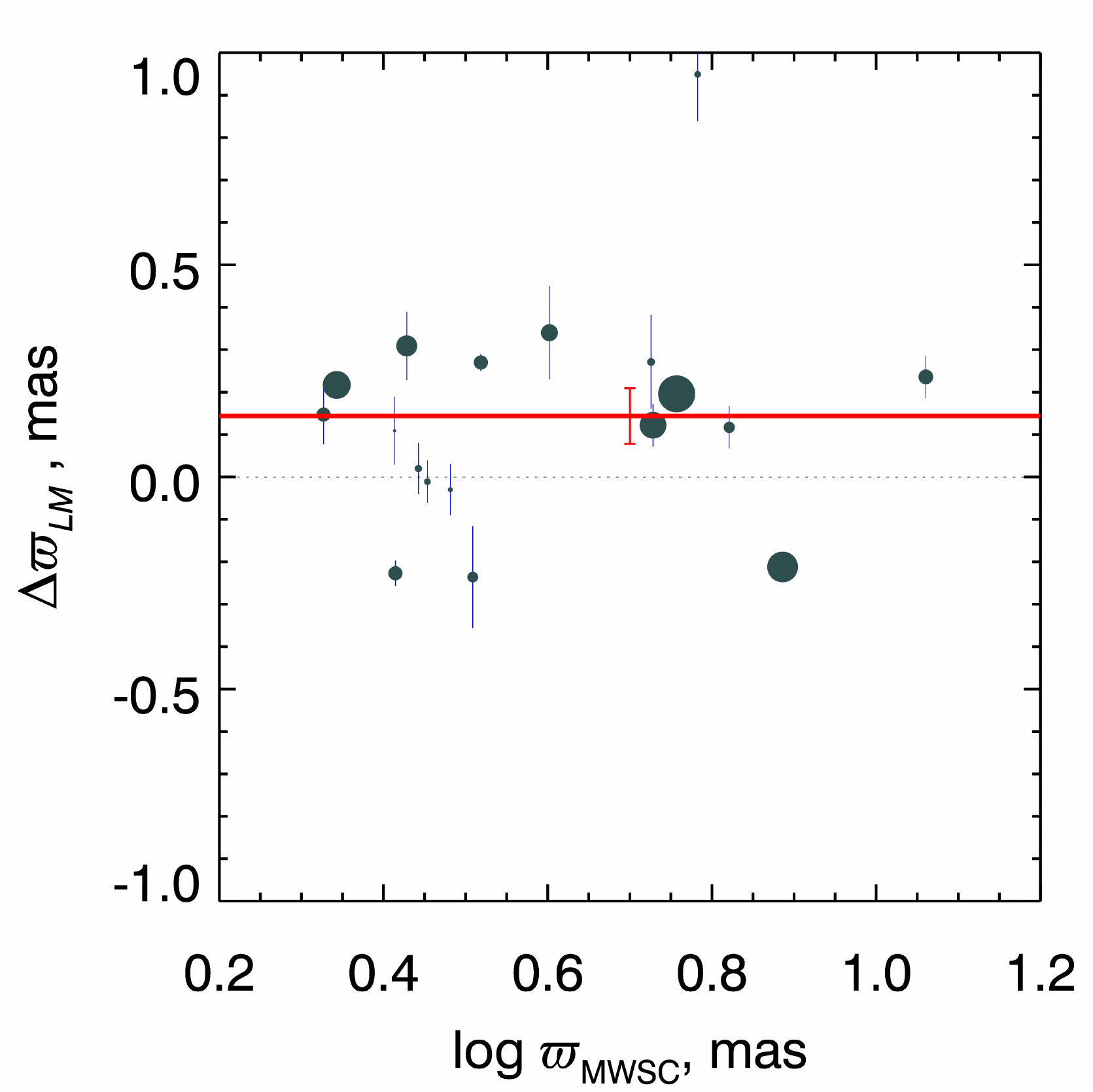}
   \includegraphics[width=0.33\hsize]{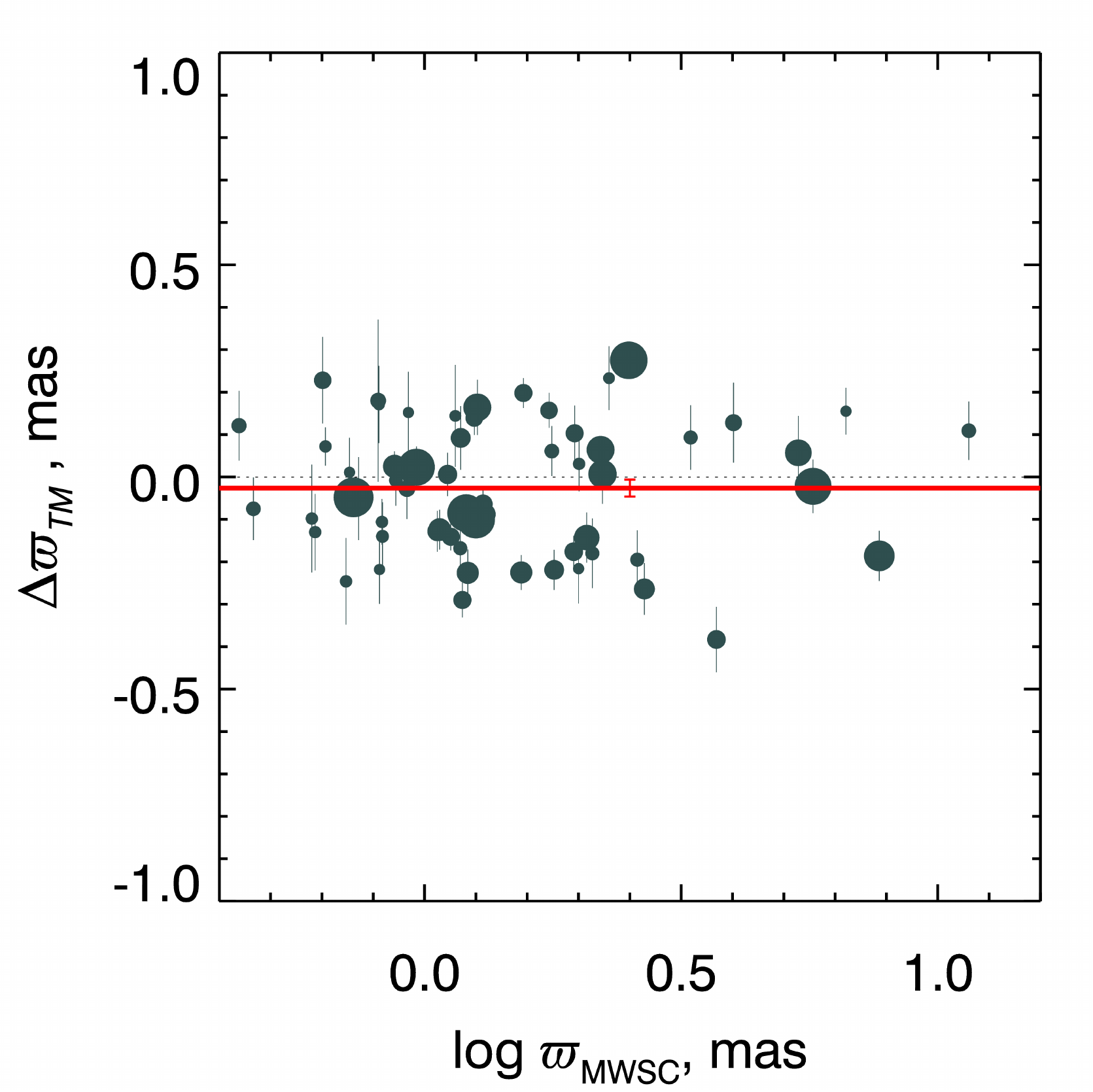}
     \caption{Differences in parallaxes  $\Delta\varpi_{LT}$, $\Delta\varpi_{LM}$,$\Delta\varpi_{TM}$ (from left to right) as a function of photometric parallax $\log \varpi_{MWSC}$. The size of the circles is proportional to the number of individual parallaxes used for the computation of the averages. Vertical bars represent the mean errors of the differences computed as described in the text. The crosses in the left panel indicate vL2017 clusters with TGAS parallaxes for $<$17 {\it probable} MWSC members. The most distant cluster used in our comparison (data point at left edge in right panel) is the rich young cluster \object{NGC 869} at about 2.3\,kpc. The thick red lines indicate the mean differences and their mean errors.}\label{fig:dplxtgas}
    \end{figure*}

We compare the trigonometric and photometric parallaxes for these clusters in Fig.~\ref{fig:avplx}. The clusters are distributed over a large range of distances up to $d_{MWSC}=2.3$\,kpc. Except for a few cases, there is good agreement between the trigonometric and photometric data, which are well concentrated to the bisector. 
A linear regression derived from the weighted least squares bisector fit \citep{1990ApJ...364..104I} with weights  computed as $1/\varepsilon_\varpi^2$ shows a well-defined linear relation
$$
\log\varpi_{TGAS} = a + b\,\log\varpi_{MWSC,}
$$
with coefficients $a=-0.00689\pm0.00002$\,mas, $b=1.01085\pm0.00002$.

In Fig.~\ref{fig:avplx}, we indicate four clusters by open circles for which the trigonometric parallaxes are significantly smaller than the photometric parallaxes. These clusters are \object{ASCC 68}, \object{Platais 12}, \object{Per OB2}, and \object{NGC 6405}. To understand the reason for this discrepancy, we looked closer into the parallax data of the {\it probable} members in these clusters. As an example, we show the case of \object{Platais 12} in Fig.~\ref{fig:plxks}. Here the parallax distribution has a bimodal character that results from a strong contamination by faint background field stars. In these few cases, a simple $3\sigma$ clipping is not an effective approach to exclude numerous non-members from the determination of the mean trigonometric parallaxes of the clusters. If we use the actual cluster members to compute the mean cluster parallax, we obtain values consistent with the corresponding photometric parallaxes. We are convinced that the uncertainty of the membership determination for faint stars in MWSC is a consequence of the low accuracy of their input data, especially of the proper motions. It is important to note that, in contrast to the mean trigonometric parallaxes, the photometric distances are only marginally affected by a poor membership determination of faint stars, since the isochrone fitting is generally controlled by bright stars that usually have more accurate input data.
       
Recently, the trigometric parallaxes of 19 selected nearby star clusters were published by \citet[][referred hereafter as \citetalias{2017A&A...601A..19G}]{2017A&A...601A..19G}. The mean cluster parallaxes are  based on a careful analysis of TGAS data. The cluster members were selected by use of position, proper motion, and parallax information.
Except for the Hyades cluster,which was not included in the  MWSC project, all other clusters have MWSC counterparts. However, eight of these contain less than the 17 {\it probable} members requested above.

In the left panel of Fig.~\ref{fig:dplxtgas}, we show the differences $\Delta\varpi_{LT}$ between the mean parallaxes derived  by \citetalias{2017A&A...601A..19G} and by us (i.e. $\varpi_{TGAS}$) as described above. The error bars stand for the mean error of the differences. The average of the differences is insignificant, i.e. $\langle\Delta\varpi_{LT}\rangle=0.10\pm0.06$\,mas. Although this is a good agreement, the \citetalias{2017A&A...601A..19G} values are not identical to ours owing to different membership criteria used by us and by  \citetalias{2017A&A...601A..19G}. We used the MWSC membership probabilities based on kinematic and photometric criteria, and on PPMXL and 2MASS data, while \citetalias{2017A&A...601A..19G} membership considered the spatio-kinematic analysis of TGAS data alone. 

The middle panel shows the differences $\Delta\varpi_{LM}$ between the trigonometric parallaxes of \citetalias{2017A&A...601A..19G} and the  MWSC photometric parallaxes $\varpi_{MWSC}$. The error bars show only the mean errors of the average parallaxes of \citetalias{2017A&A...601A..19G}. Again, the mean difference is of a low significance, i.e. $\langle\Delta\varpi_{LM}\rangle=0.14\pm0.07$\,mas. However, one observes a remarkable discrepancy between the trigonometric and photometric parallaxes of the cluster \object{IC 2391} (a dot at $\Delta\varpi_{LM}=0.95$,  $\log\varpi_{MWSC}=0.78$). To understand this disagreement, we considered all trigonometric parallax determinations for this cluster in the post-Hipparcos era, and compared these to photometric parallaxes determined at about the same time. The average trigonometric parallax of $\langle\varpi_{tr}\rangle\approx6.82\pm0.10$\,mas was computed from Hipparcos \citep{rob99,2001AstL...27..386L,2009A&A...497..209V} and TGAS data  (\citetalias{2017A&A...601A..19G}, this paper). The average photometric parallax of $\langle\varpi_{ph}\rangle\approx5.98\pm0.21$\,mas was calculated from determinations of \citet{1969AJ.....74..899P}, \citet{2001AstL...27..386L}, \citet{loktin01}, the COCD and MWSC surveys, and the data in the DAML02 database\footnote{http://www.wilton.unifei.edu.br/ocdb/}. A difference of about 1\,mas seems to be significant, so this may be a specific case in which the astrometric and photometric approaches consequently provide different  distance estimates. The cluster \object{IC 2391} is rather young (about 35 Myrs) with a rather loose colour-magnitude diagram, accompanied by a number of stars slightly below the main sequence, which mimic a larger photometric distance of the cluster. A similar effect can also occur in clusters with strongly variable extinction. Gaia DR2 will clarify the issue of \object{IC 2391}.  

The right panel is a replication of Fig.~\ref{fig:avplx} in another form but without the four discordant clusters, which we excluded from the following consideration. The error bars show only the mean errors of the average trigonometric parallaxes $\varepsilon_\varpi$. The average difference between photometric and trigonometric parallaxes $\langle\Delta\varpi_{TM}\rangle=-0.03\pm0.02$\,mas is insignificant. No systematic behaviour can be observed for the differences $\Delta\varpi_{TM}$ over the whole distance range. We conclude that, on average, the astrometric and photometric distance scales are in consistence for Galactic open clusters up to at least 2.3\,kpc from the Sun.

The estimated $\approx$10\% relative errors of the photometric
distances of MWSC clusters \citep{2013A&A...558A..53K} would
correspond to very different error bars in Fig.~\ref{fig:dplxtgas} 
(middle and right panel), rising from less 
than $\pm$0.1\,mas for distant clusters ($\log \varpi_{MWSC} < 0.0$)
to about $\pm$0.3\,mas for clusters at $\log \varpi_{MWSC} = 0.5$ and 
even about $\pm$0.5$\dots$$\pm$1\,mas for the nearest 
clusters ($\log \varpi_{MWSC} > 0.7$),
respectively. In contrast, the shown error bars of the trigonometric 
parallaxes are relatively constant at about $\pm$0.1\,mas
over all distances. As we do not observe an increase in the dispersion of 
the differences $\Delta\varpi_{LM}$ and $\Delta\varpi_{TM}$
with $\log \varpi_{MWSC}$, we conclude that the relative errors
of the photometric distances of MWSC clusters in fact decrease
with smaller cluster distances.

\citet{2016A&A...595A...4L} mentioned that the TGAS parallaxes may show systematic errors depending, for example, on positions on the sky and/or colour of the stars at a level of $\pm 0.3$\,mas. From the comparison with the Hipparcos astrometric data for open clusters, \citetalias{2017A&A...601A..19G} found that the TGAS and Hipparcos parallaxes are consistent with a slightly lower level of systematics, at 0.25\,mas. From the comparison of the TGAS trigonometric parallaxes with MWSC photometric parallaxes (see Fig.~\ref{fig:dplxtgas}, right panel), the absolute parallax differences are distributed with a standard deviation of 0.12\,mas and they are practically always smaller than 0.3\,mas. Moreover, we do not reveal a significant systematics as a function of the distance of the clusters. 

\begin{figure}
   \centering
   \includegraphics[width=\hsize]{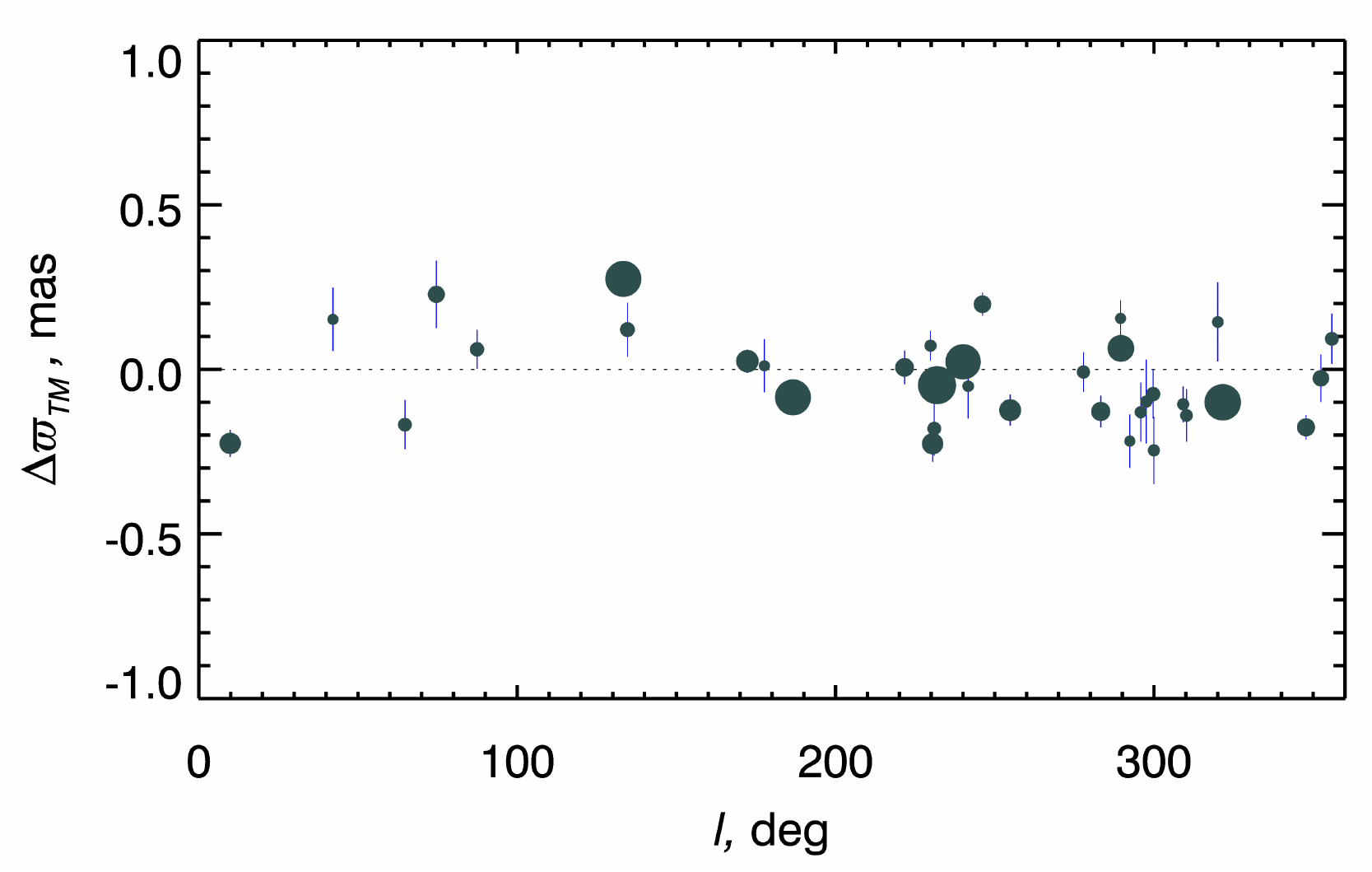}
      \caption{Differences of average trigonometric and photometric parallaxes of clusters $\Delta\varpi_{TM}$ versus galactic longitude $l$. The size of the circles is proportional to the number of TGAS parallaxes used for the calculation of averages; the vertical bars represent the error of the average TGAS parallax. }
         \label{fig:dpigl}
   \end{figure}

As the population of open clusters is strongly concentrated to the Galactic plane, we can go one step further and check for systematics within a strip of 10 degree width along the Galactic equator. Fig. \ref{fig:dpigl} shows the result for 36 clusters of our sample that are located in this strip. From this distribution one gets a hint of some systematic variations of the differences between TGAS and MWSC as a function of galactic longitude. Whereas the parallax differences show a positive offset of $\langle\Delta\varpi_{TM}\rangle=+0.11\pm0.06$\,mas at $30^\circ < l < 160^\circ$, they are, on average, smaller elsewhere, i.e. $\langle\Delta\varpi_{TM}\rangle=-0.06\pm0.02$\,mas. As the method of isochrone fitting to get photometric distances is independent of the position on the sky, it is possible that along the Galactic equator the TGAS parallaxes have  small regional systematic variations.

\section{Summary and conclusions}\label{sec:concl}
We reported on our comparison of trigonometric and photometric distance scales using open clusters. Such a comparison became possible  after the release of the Gaia DR1/TGAS catalogue. We determined the trigonometric parallaxes of MWSC clusters as average parallaxes of {\it probable} MWSC cluster members that are in TGAS, while the photometric parallaxes are taken from the published distances in MWSC, which are derived from isochrone fitting. For the calculation of $3\sigma$ clipped average TGAS parallaxes, we used the MWSC information on kinematic-photometric cluster membership. 

To compare both scales we used 64 clusters within about 2.3\,kpc from the Sun, of which each had TGAS parallax measurements of $>$16 {\it probable} members. Our major conclusion is that there is a good agreement between the trigonometric TGAS-based and the photometric MWSC-based distance scales of open clusters, which for distances of less  than 2.3\,kpc coincide at a level of about 0.1\,mas. In other words, the evolutionary stellar models that we used for isochrone fitting are sufficiently well calibrated (at least for the 2MASS passbands and stellar masses down to $0.5 M_{\odot}$) and provide unbiased distances to open clusters.  
In the coming years the measurements by Gaia \citep{2016A&A...595A...1G} will revolutionise the distance scale in our Galaxy by providing highly accurate data on trigonometric parallaxes of stars with distances up to several kpc from the Sun.  However, for much more remote clusters we will still be obliged to use isochrone fitting for distance estimates in the future. 
\begin{acknowledgements}
This study was supported by Sonderforschungsbereich SFB 881 “The Milky Way System” (subproject B5) of the German Research Foundation (DFG). Part of this work was supported by the Russian Federation of Basic Researches project number 16-52-12027. This work has made use of data from the European Space Agency (ESA) mission {\it Gaia} (\url{https://www.cosmos.esa.int/gaia}), processed by the {\it Gaia} Data Processing and Analysis Consortium (DPAC,
\url{https://www.cosmos.esa.int/web/gaia/dpac/consortium}). Funding for the DPAC has been provided by national institutions, in particular the institutions participating in the {\it Gaia} Multilateral Agreement. 
We thank the referee for comments that helped us improve the paper.
\end{acknowledgements}
%
%
\bibliographystyle{aa}
\bibliography{dk}

\begin{thebibliography}{19}
\expandafter\ifx\csname natexlab\endcsname\relax\def\natexlab#1{#1}\fi

\bibitem[{{Arenou} {et~al.}(2017){Arenou}, {Luri}, {Babusiaux}, {Fabricius},
  {Helmi}, {Robin}, {Vallenari}, {Blanco-Cuaresma}, {Cantat-Gaudin},
  {Findeisen}, {Reyl{\'e}}, {Ruiz-Dern}, {Sordo}, {Turon}, {Walton}, {Shih},
  {Antiche}, {Barache}, {Barros}, {Breddels}, {Carrasco}, {Costigan},
  {Diakit{\'e}}, {Eyer}, {Figueras}, {Galluccio}, {Heu}, {Jordi},
  {Krone-Martins}, {Lallement}, {Lambert}, {Leclerc}, {Marrese}, {Moitinho},
  {Mor}, {Romero-G{\'o}mez}, {Sartoretti}, {Soria}, {Soubiran}, {Souchay},
  {Veljanoski}, {Ziaeepour}, {Giuffrida}, {Pancino}, \&
  {Bragaglia}}]{2017A&A...599A..50A}
{Arenou}, F., {Luri}, X., {Babusiaux}, C., {et~al.} 2017, \aap, 599, A50

\bibitem[{{Gaia Collaboration} {et~al.}(2016){Gaia Collaboration}, {Prusti},
  {de Bruijne}, {Brown}, {Vallenari}, {Babusiaux}, {Bailer-Jones}, {Bastian},
  {Biermann}, {Evans}, \& et~al.}]{2016A&A...595A...1G}
{Gaia Collaboration}, {Prusti}, T., {de Bruijne}, J.~H.~J., {et~al.} 2016,
  \aap, 595, A1

\bibitem[{{Gaia Collaboration} {et~al.}(2017){Gaia Collaboration}, {van
  Leeuwen}, {Vallenari}, {Jordi}, {Lindegren}, {Bastian}, {Prusti}, {de
  Bruijne}, {Brown}, {Babusiaux}, \& et~al.}]{2017A&A...601A..19G}
{Gaia Collaboration}, {van Leeuwen}, F., {Vallenari}, A., {et~al.} 2017, \aap,
  601, A19, (vL2017)

\bibitem[{{H{\o}g} {et~al.}(2000){H{\o}g}, {Fabricius}, {Makarov}, {Urban},
  {Corbin}, {Wycoff}, {Bastian}, {Schwekendiek}, \& {Wicenec}}]{tycho2}
{H{\o}g}, E., {Fabricius}, C., {Makarov}, V.~V., {et~al.} 2000, \aap, 355, L27

\bibitem[{{Isobe} {et~al.}(1990){Isobe}, {Feigelson}, {Akritas}, \&
  {Babu}}]{1990ApJ...364..104I}
{Isobe}, T., {Feigelson}, E.~D., {Akritas}, M.~G., \& {Babu}, G.~J. 1990, \apj,
  364, 104

\bibitem[{{Kharchenko} {et~al.}(2012){Kharchenko}, {Piskunov}, {Schilbach},
  {R{\"o}ser}, \& {Scholz}}]{2012A&A...543A.156K}
{Kharchenko}, N.~V., {Piskunov}, A.~E., {Schilbach}, E., {R{\"o}ser}, S., \&
  {Scholz}, R.-D. 2012, \aap, 543, A156

\bibitem[{{Kharchenko} {et~al.}(2013){Kharchenko}, {Piskunov}, {Schilbach},
  {R{\"o}ser}, \& {Scholz}}]{2013A&A...558A..53K}
{Kharchenko}, N.~V., {Piskunov}, A.~E., {Schilbach}, E., {R{\"o}ser}, S., \&
  {Scholz}, R.-D. 2013, \aap, 558, A53

\bibitem[{{Lindegren} {et~al.}(2016){Lindegren}, {Lammers}, {Bastian},
  {Hern{\'a}ndez}, {Klioner}, {Hobbs}, {Bombrun}, {Michalik}, {Ramos-Lerate},
  {Butkevich}, {Comoretto}, {Joliet}, {Holl}, {Hutton}, {Parsons},
  {Steidelm{\"u}ller}, {Abbas}, {Altmann}, {Andrei}, {Anton}, {Bach},
  {Barache}, {Becciani}, {Berthier}, {Bianchi}, {Biermann}, {Bouquillon},
  {Bourda}, {Br{\"u}semeister}, {Bucciarelli}, {Busonero}, {Carlucci},
  {Casta{\~n}eda}, {Charlot}, {Clotet}, {Crosta}, {Davidson}, {de Felice},
  {Drimmel}, {Fabricius}, {Fienga}, {Figueras}, {Fraile}, {Gai}, {Garralda},
  {Geyer}, {Gonz{\'a}lez-Vidal}, {Guerra}, {Hambly}, {Hauser}, {Jordan},
  {Lattanzi}, {Lenhardt}, {Liao}, {L{\"o}ffler}, {McMillan}, {Mignard}, {Mora},
  {Morbidelli}, {Portell}, {Riva}, {Sarasso}, {Serraller}, {Siddiqui}, {Smart},
  {Spagna}, {Stampa}, {Steele}, {Taris}, {Torra}, {van Reeven}, {Vecchiato},
  {Zschocke}, {de Bruijne}, {Gracia}, {Raison}, {Lister}, {Marchant},
  {Messineo}, {Soffel}, {Osorio}, {de Torres}, \&
  {O'Mullane}}]{2016A&A...595A...4L}
{Lindegren}, L., {Lammers}, U., {Bastian}, U., {et~al.} 2016, \aap, 595, A4

\bibitem[{{Loktin} \& {Beshenov}(2001)}]{2001AstL...27..386L}
{Loktin}, A.~V. \& {Beshenov}, G.~V. 2001, Astronomy Letters, 27, 386

\bibitem[{{Loktin} {et~al.}(2001){Loktin}, {Gerasimenko}, \&
  {Malysheva}}]{loktin01}
{Loktin}, A.~V., {Gerasimenko}, T.~P., \& {Malysheva}, L.~K. 2001, Astronomical
  and Astrophysical Transactions, 20, 607

\bibitem[{{Michalik} {et~al.}(2015){Michalik}, {Lindegren}, \&
  {Hobbs}}]{2015A&A...574A.115M}
{Michalik}, D., {Lindegren}, L., \& {Hobbs}, D. 2015, \aap, 574, A115

\bibitem[{{Perry} \& {Hill}(1969)}]{1969AJ.....74..899P}
{Perry}, C.~L. \& {Hill}, G. 1969, \aj, 74, 899

\bibitem[{{Robichon} {et~al.}(1999){Robichon}, {Arenou}, {Mermilliod}, \&
  {Turon}}]{rob99}
{Robichon}, N., {Arenou}, F., {Mermilliod}, J.-C., \& {Turon}, C. 1999, \aap,
  345, 471

\bibitem[{{R{\"o}ser} {et~al.}(2010){R{\"o}ser}, {Demleitner}, \&
  {Schilbach}}]{2010AJ....139.2440R}
{R{\"o}ser}, S., {Demleitner}, M., \& {Schilbach}, E. 2010, \aj, 139, 2440

\bibitem[{{Schmeja} {et~al.}(2014){Schmeja}, {Kharchenko}, {Piskunov},
  {R{\"o}ser}, {Schilbach}, {Froebrich}, \& {Scholz}}]{2014A&A...568A..51S}
{Schmeja}, S., {Kharchenko}, N.~V., {Piskunov}, A.~E., {et~al.} 2014, \aap,
  568, A51

\bibitem[{{Scholz} {et~al.}(2015){Scholz}, {Kharchenko}, {Piskunov},
  {R{\"o}ser}, \& {Schilbach}}]{2015A&A...581A..39S}
{Scholz}, R.-D., {Kharchenko}, N.~V., {Piskunov}, A.~E., {R{\"o}ser}, S., \&
  {Schilbach}, E. 2015, \aap, 581, A39

\bibitem[{{Skrutskie} {et~al.}(2006){Skrutskie}, {Cutri}, {Stiening},
  {Weinberg}, {Schneider}, {Carpenter}, {Beichman}, {Capps}, {Chester},
  {Elias}, {Huchra}, {Liebert}, {Lonsdale}, {Monet}, {Price}, {Seitzer},
  {Jarrett}, {Kirkpatrick}, {Gizis}, {Howard}, {Evans}, {Fowler}, {Fullmer},
  {Hurt}, {Light}, {Kopan}, {Marsh}, {McCallon}, {Tam}, {Van Dyk}, \&
  {Wheelock}}]{cat2MASS}
{Skrutskie}, M.~F., {Cutri}, R.~M., {Stiening}, R., {et~al.} 2006, \aj, 131,
  1163

\bibitem[{{Taylor}(2005)}]{2005ASPC..347...29T}
{Taylor}, M.~B. 2005, in Astronomical Society of the Pacific Conference Series,
  Vol. 347, Astronomical Data Analysis Software and Systems XIV, ed.
  P.~{Shopbell}, M.~{Britton}, \& R.~{Ebert}, 29

\bibitem[{{van Leeuwen}(2009)}]{2009A&A...497..209V}
{van Leeuwen}, F. 2009, \aap, 497, 209

\end{thebibliography}
\end{document}